\documentclass[final]{svjour2}
\usepackage{graphicx}
\usepackage{rotating}
\usepackage{amssymb}
\usepackage{mathptmx}
\usepackage[numbers]{natbib}
\usepackage{bm}
\makeatletter\journalname{Journal of Low Temperature Physics}

\bibpunct{}{}{,}{s}{}{,}

\begin{document}

\newcommand{\hdblarrow}{H\makebox[0.9ex][l]{$\downdownarrows$}-}
\title{Decay of Counterf\mbox{}low Quantum Turbulence in Superf\mbox{}luid $^4$He}

\author{Y. Mineda$^1$ \and M. Tsubota$^{1,2}$ \and W.F. Vinen$^3$}
\institute{1:Department of Physics, Osaka City University,\\ Sumiyoshi-Ku,Osaka 558-8585, Japan\\
Tel.: +81-6-6605-2504\\ Fax: +81-6-6605-2522\\
\email{mineda@sci.osaka-cu.ac.jp}
\\2:The OCU Advanced Research Institute for Natural Science and Technology (OCARINA), Osaka City University,\\ Sumiyoshi-Ku, Osaka 558-8585, Japan
\\3:School of Physics and Astronomy, University of Birmingham B15 2TT, United Kingdom\\}

\date{\today}

\maketitle

\keywords{$^4$He, quantized vortices, thermal counterf\mbox{}low, decay turbulence}
\begin{abstract}We have simulated the decay of thermal counterf\mbox{}low quantum turbulence from a statistically steady state at $T$=1.9[K], with the assumption that the normal f\mbox{}luid is at rest during the decay.
The results are consistent with the predictions of the Vinen equation (in essence the vortex line density (VLD) decays as $t^{-1}$).
For the statistically steady state, we determine the parameter $c_2$, which connects the curvature of the vortex lines and the mean separation of vortices.
A formula connecting the parameter $\chi_2$ of the Vinen equation with $c_2$ is shown to agree with the results of the simulations.
Disagreement with experiment is discussed.

PACS numbers: 67.25.dk,67.25.dg
\end{abstract}

\section{Introduction}
Superf\mbox{}luid $^4$He at a f\mbox{}inite temperature behaves as an an intimate mixture of two components: an inviscid superf\mbox{}luid component (density $\rho_{{\mathrm{s}}}$ and velocity $\bm v_{\mathrm{s}}$); and a viscous normal component (density $\rho_{{\mathrm{n}}}$ and velocity $\bm v_{\mathrm{n}}$).
Quantum turbulence\cite{Hal,Tsu} takes the form of a tangle of quantized vortices in the superf\mbox{}luid component; the normal f\mbox{}luid may or may not be turbulent.
Thermal counterf\mbox{}low is an internal convection in which the superf\mbox{}luid component f\mbox{}lows towards a source of heat, while the normal f\mbox{}luid, carrying all the entropy, f\mbox{}lows away from it.
There is a relative velocity $\bm v_{\mathrm{ns}}=\bm v_{\mathrm{n}}-\bm v_{\mathrm{s}}$ between the two f\mbox{}luids, with no net mass f\mbox{}low.
When the counterf\mbox{}low velocity exceeds a certain critical value, laminar f\mbox{}low proves to be unstable and the superf\mbox{}luid component becomes turbulent.
This was the f\mbox{}irst type of quantum turbulence to be identif\mbox{}ied experimentally in the late 1950s\cite{Vin1}. 

The vortices in quantum turbulence interact with the normal f\mbox{}luid, giving rise a force of "mutual friction" between the two f\mbox{}luids.
The study of the characteristics of this mutual friction provided the experimental evidence for counterf\mbox{}low turbulence.
The turbulence in the superf\mbox{}luid component can be characterized by the vortex line density (VLD), $L(t)$, and it was suggested that the time-evolution of this density in thermal counterf\mbox{}low might be described by the (Vinen) equation
\begin{equation}
\frac{dL}{dt}=\alpha\left|\bm v_{\mathrm{ns}}\right|L^{3/2}-\chi_2\frac{\kappa}{2\pi}L^2,
\label{eq:V_eq}
\end{equation}
where $\alpha$ and $\chi_2$ are temperature-dependent parameters and $\kappa$ is the quantum of circulation.
The f\mbox{}irst term represents the generation of counterf\mbox{}low turbulence and the second term represents its decay.
This equation implies that counterf\mbox{}low turbulence is homogeneous, which appears indeed to be the case as long as the spacing between vortices is much smaller than the size of the channel in which the heat current f\mbox{}lows. 

Since the 1950s, many experimental, theoretical and numerical studies of coun- terf\mbox{}low turbulence have been reported.
Schwarz showed by numerical simulations with the vortex f\mbox{}ilament model\cite{Sch1} that a self-sustained statistically steady-state form of turbulence can be maintained in the superf\mbox{}luid component\cite{Sch2} by the mutual friction, and a more satisfactory simulation based on a full Biot-Savart simulation has recently been performed by Adachi {\it et al.}\cite{Ada}.
In both simulations f\mbox{}low of the normal f\mbox{}luid is assumed to be laminar.
The steady-state turbulence is realized when the generation and decay terms in Eq. (\ref{eq:V_eq}) balance, so that $L=L_{\mathrm{steady}}=\gamma^2v_{\mathrm{ns}}^2$, where $\gamma$ is a temperature-dependent parameter. 

Counterf\mbox{}low turbulence in the steady state seems now to be well understood, but its decay, after the heat current is switched off, is less well understood.
According to Eq.(\ref{eq:V_eq}), with $v_{\mathrm{ns}}=0$, this decay should obey the equation
\begin{equation}
\frac{1}{L}=\chi_2\frac{\kappa}{2\pi}t+\frac{1}{L_0},
\label{eq:V_sol}
\end{equation}
where $L_0$ is the initial VLD.
However, it has been known for a long time that in practice the decay proceeds in a more complicated (non-monotonic) manner\cite{Vin1}, although there is some evidence that Eq. (\ref{eq:V_sol}) is obeyed at \textit{small} times.
It seems now to be well established\cite{Skr} that for \textit{large} times the decay proceeds as $t^{-3/2}$, implying that at this stage there is coupled motion of the two f\mbox{}luids dominated by eddies on a scale equal to the channel width\cite{Sta}; i.e. the normal f\mbox{}luid is also turbulent. 

 In this paper we report the results of simulations of the decay of counterf\mbox{}low turbulence, starting from a statistically steady state, in order to check the predictions of Eq. (\ref{eq:V_eq}).
We assume that the normal f\mbox{}luid remains at rest during the decay.
The validity of our simulations is limited to the extent that our VLDs are somewhat smaller than those encountered in practice, larger VLDs being hard to handle, but within this limitation our results prove to be consistent with the predictions of Eq. (\ref{eq:V_eq}), with the expected value of $\chi_2$.
  
\section{The simulation}
We use the vortex f\mbox{}ilament model\cite{Sch1,Sch2} and represent a vortex as a string of discrete points ${\bm s}\left( \xi, t \right)$, where $t$ is time and $\xi$ is arc length.
At 0 K the velocity $\dot {\bm s_{0}}$ of the f\mbox{}ilament at the point ${\bm s}$ is given by
\begin{equation}
\dot {\bm s_{0}}=\frac{\kappa}{4\pi}{\bm s}' \times {\bm s}'' \ln \left( \frac{2\left( l_{+}l_{-}\right)^{1/2}}{e^{1/2} \xi_{0}} \right)+\frac{\kappa}{4\pi} \int _{\cal L}' \frac{\left({\bm s}-{\bm r}\right) \times d{\bm s}}{\left| {\bm s}-{\bm r} \right|^{3}}+\bm v_{\mathrm s},
\label{eq:local non-local}
\end{equation}
where the prime denotes differentiation with respect to arc length $\xi$, $l_{+}$ and $l_{-}$ are the lengths of the two adjacent line elements after discretization, separated by the point ${\bm s}$, and $\xi_{0}$ is the cutoff corresponding to the core radius.
The f\mbox{}irst term shows the localized induction f\mbox{}ield arising from a curved line element acting on itself.
The second term represents the non-local f\mbox{}ield obtained by carrying out the integral of the Biot-Savart law along the rest of the f\mbox{}ilament.
The third term $\bm v_{s}$ is an applied f\mbox{}ield. 

At f\mbox{}inite temperatures, it is necessary to take into account of the mutual friction between the vortex core and the normal f\mbox{}low ${\bm v}_{{\mathrm n}}$, so that the velocity $\dot {\bm s}$ becomes
\begin{equation}
\dot {\bm s}=\dot {\bm s}_{0}+\alpha {\bm s}' \times \left( {\bm v}_{{\mathrm n}}-\dot {\bm s}_{0}\right)-\alpha' {\bm s}' \times \left[ {\bm s}' \times \left( {\bm v}_{{\mathrm n}}- \dot {\bm s}_{0} \right) \right],
\label{eq:vortex motion}
\end{equation}
where $\alpha$ and $\alpha'$ are temperature-dependent friction coeff\mbox{}icients\cite{Sch1}, and $\dot {\bm s}_{0}$ is calculated from Eq. (\ref{eq:local non-local}). 

Some important quantities that are useful for characterizing the vortex tangle can introduced\cite{Sch2}.
The VLD is
\begin{equation}
L=\frac{1}{\mathrm\Omega}\int_{\cal L}d\xi,
\label{eq:VLD}
\end{equation}
where the integral is performed along all vortices in the sample volume $\mathrm\Omega$.
In the experiments\cite{Vin1,Skr}, the VLD was obtained from the attenuation of second sound propagating perpendicular to the counterf\mbox{}low $\bm v_{\mathrm{ns}}$.
Because mutual friction acts only at right angles to a vortex line element, the directly measured quantity is not the real VLD given by Eq. (\ref{eq:VLD}), but rather a "measured VLD",  given by
\begin{equation}L_p=\frac{1}
{\mathrm\Omega}\int_{\cal L}\mathrm{sin}^2\theta d\xi,
\label{eq:projection VLD_xz}
\end{equation}
where $\theta$ is the angle between a vortex line segment and and the direction $\hat{\bm r}$ of propagation of the second sound.
If the quantum turbulence is isotropic, $L_p=(2/3)L$.
Anisotropy of the vortex tangle is also interesting and is described by the dimensionless parameters
\begin{eqnarray}
I_{\parallel}&=&\frac{1}{\Omega L}\int_{\cal L}\left[1-\left(\bm s'\cdot\hat{\bm r}_{\parallel}\right)^2\right]d\xi,
\label{eq:Ipa} 
\\I_{\perp}&=&\frac{1}{\Omega L}\int_{\cal L}\left[1-\left(\bm s'\cdot\hat{\bm r}_{\perp}\right)^2\right]d\xi,
\label{eq:Ino}
\end{eqnarray}
where $\hat{\bm r}_{\parallel}$ and $\hat{\bm r}_{\perp}$ represent unit vectors parallel and perpendicular to the original heat f\mbox{}low (in our case $\hat{\bm r}_{\parallel}=\hat{\bm z}$).
Since $\hat{\bm r}_{\perp}$ is the same as the direction of propagation of the second sound, then $L_p/L=I_{\perp}$.
Symmetry generally yields the relation $I_{\parallel}/2+I_{\perp}=1$.
If the vortex tangle is isotropic, the averages of these parameters are $\overline{I}_{\parallel}=\overline{I}_{\perp}=2/3$.
At the other extreme, if the tangle consists entirely of curves lying in planes normal to $\bm v_{\mathrm{ns}}$, $\overline{I}_{\parallel}=1$ and $\overline{I}_{\perp}=1/2$.

In this study, calculations are performed under the following conditions.
The numerical space resolution is $\Delta\xi=8.0\times10^{-4}$[cm], the time resolution is $\Delta t=1.0\times10^{-4}$[s], temperature is 1.9[K], and the computing box size is $0.1\times0.1\times0.1$[cm$^3$].
We use periodic conditions in all directions.

\section{Results}
\subsection{Turbulence in the statistically steady state}
Our decays start from a statistically steady state, and it is instructive f\mbox{}irst to describe the character of this state and how it is generated\cite{Ada}. 

First, we place in the box six vortex rings moving towards its centre, and we impose the relative velocity, $v_{\mathrm{ns}}=0.52$[cm/s], along the $z$-axis at $T$=1.9[K].
The vortices make many reconnections and the VLD increases until it becomes statistically steady.
The VLD of this steady state turbulence is $L_0\sim6.04\times10^3[1/\mathrm{cm}^2]$ and it satisf\mbox{}ies the equation $L=\gamma^2v_{\mathrm{ns}}^2$. 

We investigated the relationship\cite{Sch2} between the curvature of the vortex lines and the mean distance between vortices, $l=1/L^{1/2}$,
\begin{equation}\left<\frac{1}{R^2}\right>=c_2^2\frac{1}{l^2},
\label{eq:c2}
\end{equation}
where the left hand side is the mean square curvature of the vortex lines and $c_2$ is a temperature-dependent parameter.
The parameter $c_2$ tends to decrease with increasing temperature, because the high velocities associated with small radii of curvature are more strongly damped by mutual friction at high temperature. 

\begin{figure}[htbp]
\begin{center}
\begin{minipage}{.49\linewidth}(a) \\\includegraphics[width=\linewidth,keepaspectratio]{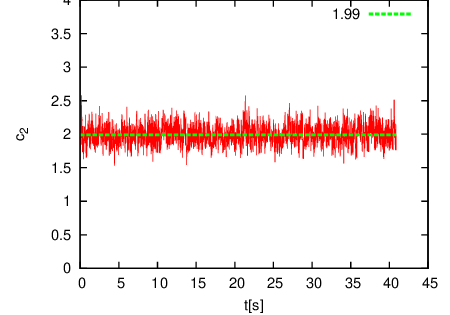}
\end{minipage}
\begin{minipage}{.49\linewidth}(b) \\\includegraphics[width=\linewidth,keepaspectratio]{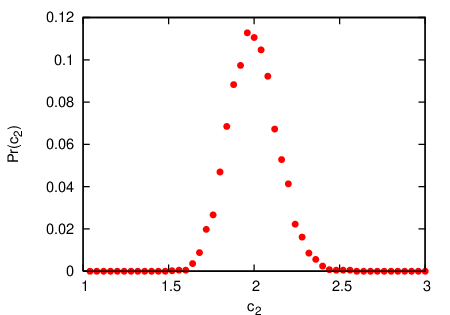}
\end{minipage}
\caption{(Color online) (a) Dependence of $c_2$ on time in the statistically steady state. The parameter $c_2$ f\mbox{}luctuates around 1.99. (b) Distribution of $c_2$. The peak value is $c_2=1.96$ and the standard deviation is $\sigma_{c_2}=0.38$.}
\label{c2}
\end{center}
\end{figure}
Figure \ref{c2} (a) shows that the parameter $c_2$ is statistically constant and the time average of $c_2$ is 1.99.
Apparently, the f\mbox{}luctuation of $c_2$ comes from the f\mbox{}luctuations in the statistically steady state of the VLD  and the mean square curvature.
We plot the probability density distribution of $c_2$ in Fig. \ref{c2} (b) obtaining $c_2=1.99\pm 0.38$, where the error is a standard deviation.
The dependence of $c_2$ on temperature will be reported in another paper.

\subsection{Decay of the counterf\mbox{}low turbulence}
\begin{figure}[htbp]
\begin{center}
\begin{minipage}{.49\linewidth}(a)@\\\includegraphics[width=\linewidth,keepaspectratio]{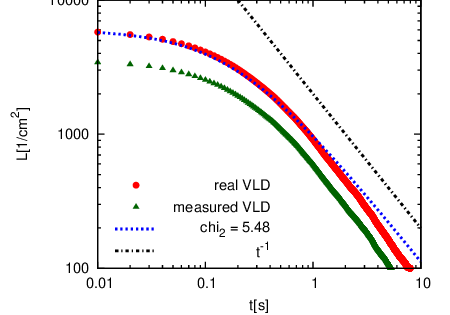}
\end{minipage}
\begin{minipage}{.49\linewidth}(b)  \\\includegraphics[width=\linewidth,keepaspectratio]{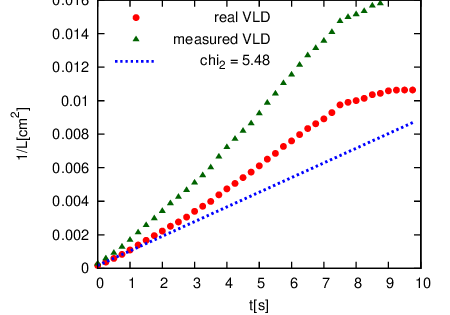}
\end{minipage}
\caption{(Color online) The decay of VLD (red circle points) and the measured VLD (green triangle points). These f\mbox{}igures are averaged data. Figure (a) is a log-log plot of $L$ against $t$,  and Fig. (b) is a normal plot of $1/L$ against $t$. The blue dotted line is the solution of Vinen equation ($\chi_2=5.48$). The black dot-dashed line shows $t^{-1}$ power law.}
\label{T19VLD}
\end{center}
\end{figure}

After we obtained the steady state, we turned off the relative velocity, so that the turbulence can decay.
It is necessary to study the statistics of the decaying turbulence, because the relatively small initial VLDs lead to f\mbox{}luctuations of the decaying VLD that are not  negligible in one decay.
Therefore we simulated 41 decays from slightly different initial conditions, and then averaged the results.
Figure \ref{T19VLD} shows these averaged results, in the form of plots of $L$ and $1/L$ against time.
Red circle points are the real VLDs and green triangle points show how the "measured VLDs" are expected to behave.

\begin{figure}[htbp]
\begin{center}
\includegraphics[width=\linewidth,keepaspectratio]{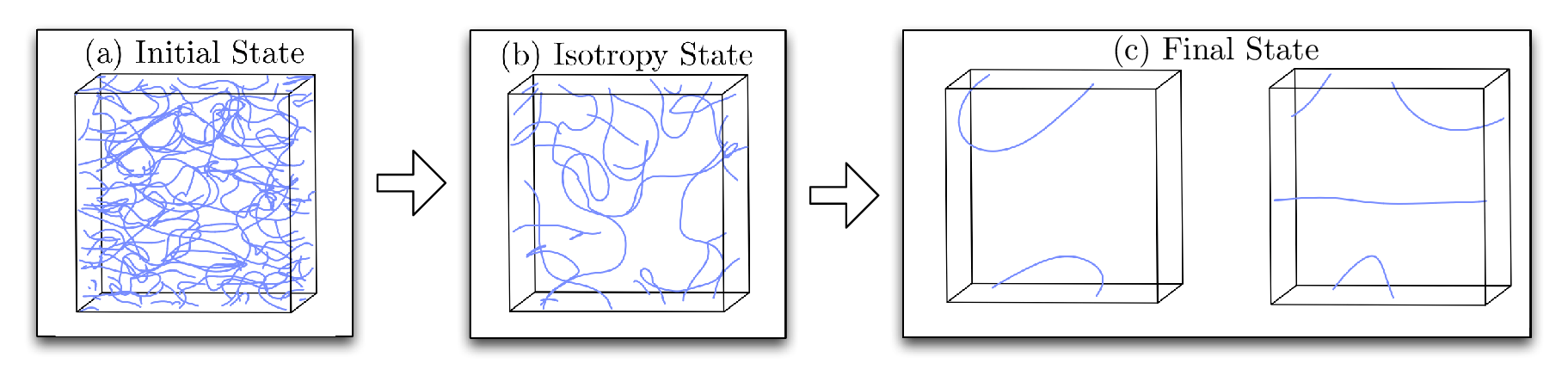}
\caption{(Color online) Conf\mbox{}iguration of the vortex tangle at (a) $t$=0.0[s],  (b) $t\sim$0.5[s],  and  (c) $t>$3.0[s].}
\label{decay}
\end{center}
\end{figure}

Figure \ref{decay} shows the conf\mbox{}iguration of vortices in the decaying turbulence at various times.
As we demonstrate below, the conf\mbox{}iguration is initially anisotropic (a), but after about 0.5[s] it becomes isotropic (b), until at times greater than about 3[s] the line density becomes so small (line spacing comparable with the size of the box) that the decay is strongly inf\mbox{}luenced by the walls of the box. 

We see from Fig. \ref{T19VLD} that an equation of the form of Eq. (\ref{eq:V_sol}) is obeyed as long as the VLD is not so small that the decay is inf\mbox{}luenced by the walls of the box.
This result applies to both the real VLD and the measured VLD.

It was argued in reference\cite{Vin2} from a consideration of the total rate of energy dissipation due to mutual friction that $\chi_2$ and $c_2$ can be expected to be related by the equation
\begin{equation}\chi_2=\frac{\alpha c_2^2}{2}\ln{\frac{R_0}{\xi_0}},
\label{eq:chi2}
\end{equation}
where $R_0=1/(L_0c_2^2)$ and $L_0$ is an initial VLD.
In this way we can predict that at 1.9[K] $\chi_2=5.48\pm0.88$.
We see from Fig. \ref{T19VLD} that our simulations support this prediction. 

\begin{figure}[htbp]
\begin{center}
\begin{minipage}{.49\linewidth}(a) \\\includegraphics[width=\linewidth,keepaspectratio]{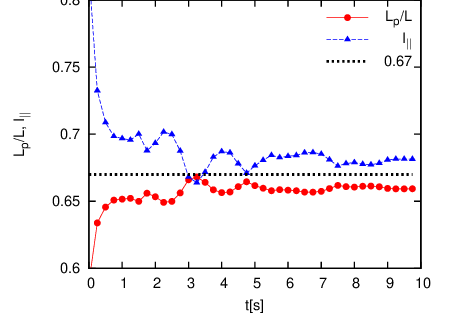}
\end{minipage}
\begin{minipage}{.49\linewidth}(b) \\\includegraphics[width=\linewidth,keepaspectratio]{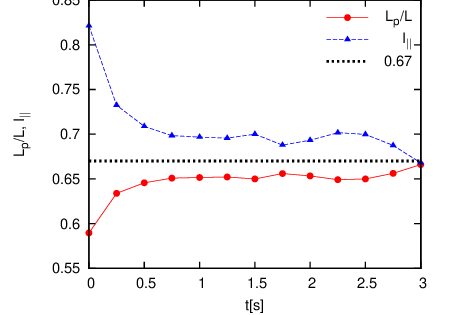}
\end{minipage}
\caption{(Color online) The ratio of $L_p$ to $L$ (red circle points) and the anisotropy parameter, $I_{\parallel}$, (blue triangle points), plotted against time. The black dotted lines shows 2/3.}
\label{Lp_L}
\end{center}
\end{figure}

The way in which the degree of anisotropy changes during the decay is shown in Fig. \ref{Lp_L}, in which we have plotted the ratio $L_p/L$ and the parameter $I_{\parallel}$ against time.
For isotropic turbulence $L_p/L=I_{\parallel}=2/3$. 

As we have ready noted, the decay at times greater than $\sim$3[s] is strongly inf\mbox{}luenced by the walls of the box, and this can be seen in the behavior of $L_p/L$ in Fig. \ref{Lp_L} (a).
Figure \ref{Lp_L} (b) shows that the decaying turbulence remains strongly anisotropic until $t\sim$0.5[s].
In the interval between $t$=0.5[s] and $t$=2.5[s], the averages of $L_p/L$ and $I_{\parallel}$ are $0.65\pm 0.04$ and $0.70\pm 0.08$, both values being close to 2/3, indicating that the turbulence has become approximately isotopic. 

\section{Conclusions}
We have carried out computer simulations of the decay of quantum turbulence, using the vortex f\mbox{}ilament model\cite{Sch1,Sch2}, with the assumption that the normal f\mbox{}luid is stationary during the decay.
The simulations start from a statistically steady state of thermal counterf\mbox{}low turbulence\cite{Ada} (Fig. \ref{decay} (a)),  generated by a f\mbox{}low in which $v_{\mathrm{s}}-v_{\mathrm{n}}=0.52$[cm/s] at a temperature of 1.9[K].
The results prove to be consistent with the predictions of the Vinen equation (essentially the vortex line density decays with time as $t^{-1}$ in the limit of large times, as long as the line spacing is small compared with the box size).
A parameter $c_2$, which describes the relationship between the curvature of the vortex lines and the mean distance between vortices, was evaluated for the steady state, and it is shown that the value of a parameter $\chi_2$ that appears in the Vinen equation is correctly predicted in terms of $c_2$ by a formula derived by Vinen and Niemela\cite{Vin2}.
The turbulence is markedly anisotropic in the steady state counterf\mbox{}low, but it becomes isotropic as the decay proceeds.
The validity of  these results at temperatures other than 1.9[K] will be investigated in future work. 

The type of decay that emerges from our simulations does not agree with experiment, as has been suspected for many years.
The origin of the discrepancy has been discussed by Schwarz and Rozen\cite{Sch3} and by Barenghi \textit{et al.}\cite{Bar}.
Our results do not add support to the idea of Barenghi \textit{et al.} that the discrepancy is associated in part with the transition during decay from an anisotropic to an isotropic vortex conf\mbox{}iguration.
We suspect that the discrepancy is due primarily to the existence of coupled motion of the two f\mbox{}luids during the decay (perhaps even in the steady state), as suggested  by Schwarz and Rozen. 

\begin{acknowledgements}
We would like to thank L. Skrbek and S. Babuin for helpful discussions.
\end{acknowledgements}


\end{document}